\documentclass[openacc]{rstransa}

\newcommand{\be}{\begin{equation}}
\newcommand{\ee}{\end{equation}}
\newcommand{\bq}{\begin{eqnarray}}
\newcommand{\eq}{\end{eqnarray}}

\newcommand{\dtot}[2]{\frac{d #1}{d #2}}

\begin{document}

\title{Cosmological Evolution of Semilocal String Networks}

\author{
A. Ach\'ucarro$^{1, 2}$, A. Avgoustidis$^{3}$,  \qquad\qquad A. L\'opez-Eiguren$^{4}$, C.J.A.P. Martins$^{5,6}$ and \qquad\qquad 
J. Urrestilla$^{2}$ 
}

\address{$^{1}$Institute Lorentz of Theoretical Physics, University of Leiden, 2333CA Leiden, The Netherlands\\
$^{2}$Department of Theoretical Physics, University of the Basque Country UPV-EHU, 48040 Bilbao, Spain\\
$^{3}$School of Physics and Astronomy, University of Nottingham, University Park, Nottingham NG7 2RD, England\\
$^{4}$ Department of Physics and Helsinki Institute of Physics, PL 64, FI-00014 University of Helsinki, Finland\\
$^{5}$Centro de Astrof\'{\i}sica da Universidade do Porto, Rua das Estrelas, 4150-762 Porto, Portugal\\
$^{6}$Instituto de Astrof\'{\i}sica e Ci\^encias do Espa\c co, CAUP, Rua das Estrelas, 4150-762 Porto, Portugal\\
}

\subject{field theory,  cosmology}


\corres{Ana Ach\'ucarro\\
\email{achucar@lorentz.leidenuniv.nl}}

\begin{abstract}
Semilocal strings --a particular limit of electroweak strings-- are an interesting example of a stable non-topological defect whose properties resemble those of their topological cousins, the Abrikosov-Nielsen-Olesen vortices. There is, however, one important difference: a network of semilocal strings will contain segments. These are 'dumbbells' whose ends behave almost like global monopoles that are strongly attracted to one another.  While closed loops of string will eventually shrink and disappear, the segments can either shrink or grow, and a cosmological network of semilocal strings will reach a scaling regime.   We discuss  attempts to find  a "thermodynamic" description of the cosmological evolution and scaling of a network of semilocal strings,  by analogy with well-known descriptions for cosmic strings and for monopoles. We propose a model for the time evolution of 
an overall length-scale and typical velocity for the network as well as for its segments, and some supporting (preliminary) numerical evidence.
\end{abstract}


\maketitle

\section{Introduction}

Semilocal strings are a non-topological type of defect arising in theories with both local and global symmetries 
\cite{Vachaspati:1991dz,Hindmarsh:1991jq}  (see \cite{Achucarro:1999it} for a review).
The simplest semilocal model is an  extension of the Abelian Higgs model where the Higgs scalar field is replaced by two equally charged Higgses forming an $SU(2)$ doublet.  The semilocal model can also be thought of as  the bosonic sector of the Glashow-Weinberg-Salam model of electroweak interactions, in the limit where the $SU(2)$ coupling constant is zero, the $Z $ and $W$ bosons decouple  and the $SU(2)$ symmetry becomes global. \\

The semilocal model admits stable string solutions even though the topology of the vacuum manifold would suggest otherwise
 (the vacuum manifold is a three-sphere, $S^3 $,  which is simply connected). However, being non-topological, semilocal strings have  different  properties compared to their topological counterparts in the Abelian Higgs model. The most noteworthy of these is that they can form finite open segments.
 We will return to this point below.\\
 
Semilocal string networks can arise in supersymmetric grand unified theories of inflation \cite{Urrestilla:2004eh} and in D3/D7 brane inflation models \cite{Dasgupta:2004dw}. These are a natural extension of usual inflationary models, in which the only extra ingredient is the doubling in the scalar sector. The signatures and potential detectability of a cosmological network of semilocal strings in the Cosmic Microwave Background have been considered in \cite{Urrestilla:2007sf,Sazhina:2013oua, Lizarraga:2014xza}. The latest constraints can be found in \cite{Lizarraga:2014xza,Ade:2015xua}.
\\

The model is described by the lagrangian  \cite{Vachaspati:1991dz,Achucarro:1999it}
\be
S=\int\!\! d^4x \sqrt{-g}  \!\left[ (D_\mu\Phi)^\dagger D^\mu\Phi -\frac{1}{4}F_{\mu \nu} F^{\mu \nu}
-\lambda(\Phi^+\Phi-\frac{\eta^2}{2})^2\right]\,,
\ee
where $\Phi$ is a doublet of complex scalar fields $\Phi=(\phi_1,\phi_2)$, $D_\mu\Phi=(\partial_\mu-i q A_\mu)\Phi$,   and
$F_{\mu\nu} = (\partial_\mu A_\nu - \partial_\nu A_\mu)$ is the gauge
field strength.   The model is invariant under $SU(2)_{\rm global}\times U(1)_{\rm local}$ and, after symmetry breaking, the particle content is two Goldstone bosons, one real scalar with mass $m_s=\sqrt{2\lambda}\eta$ and a vector boson with mass $m_v=q\eta$. At the classical level the dynamics is determined by a single parameter, the ratio of the scalar mass to the vector mass. This is most easily seen with the rescalings:
\begin{equation}
\Phi\to\frac{\eta}{\sqrt{2}}\Phi\,,\quad x\to \frac{\sqrt{2}}{q\eta}x\,,\quad A_\mu\to \frac{\eta}{\sqrt{2}}A_\mu\,,
\end{equation}
and introducing
\be
\beta=m_s^2/m_v^2=2\lambda  / q^2\,,
\label{beta}
\ee
the action becomes
\be
S=\int\!\! d^4x \sqrt{-g}  \!\left[ (D_\mu\Phi)^\dagger D^\mu\Phi-\frac{1}{4}F^{\mu\nu} F_{\mu\nu}
-\frac{\beta}{2}(\Phi^+\Phi-1)^2\right]\,,
\label{SLaction}
\ee
where now $D_\mu\Phi=(\partial_\mu-i A_\mu)\Phi$.  The parameter $\beta$ is analogous to the usual parameter that distinguishes Type I ($\beta<1$) from Type II ($\beta>1$) superconductors in the Abelian Higgs model. \\

The stability  of infinitely long, straight strings is not guaranteed by topology, the Higgs field can be non-zero everywhere in the core of the strings.  The magnetic flux --measured at infinity-- is quantized, but not necessariy confined to a vortex of a particular width. It turns out that the parameter $\beta$ determines the stability of the strings \cite{Hindmarsh:1991jq} : if $\beta < 1$ the strings are stable, if $\beta>1$ they are unstable. 
We can understand this intuitively as follows: if $q=0$, the theory would have a global $SU(2) \times U(1)$ symmetry, and we would expect the (global) string to be unstable. Instead, if $q$ is sufficiently large, it is energetically advantageous to maintain a region with a low Higgs vacuum expectation value. where the magnetic field can accrete. Stability is then decided by the competition of these effects. A more detailed argument \cite{Hindmarsh:1991jq} then shows that the boundary between stability and instability is at $\beta = 1$, where there is a one-parameter family of neutrally stable solutions with the same energy per unit length but with varying core widths (see also \cite{Vachaspati:1991dz}). This is usually called the BPS or supersymmetric limit. \\

Unlike Abelian Higgs strings --that have to be infinitely long or form closed loops-- semilocal strings in the stability regime $\beta<1$ can form finite segments. However, it is energetically costly to end a string and, as a result, the ends of the segments behave like global monopoles \cite{Hindmarsh:1992yy}, with long-range interactions among them.  This accounts for very rich dynamics. Closed loops behave like their topological cousins in the Abelian Higgs model, they eventually shrink and disappear. On the other hand,  segments can shrink and disappear, but they can also grow to join other segments (or to form closed loops). Modelling the cosmological evolution of such a system has proved a very non-trivial task.  \\

The results presented here are part of ongoing research by the authors, described in  \cite{Achucarro:2013mga,Nunes:2011sf,Lopez-Eiguren:2016jsy, Lopez-Eiguren:2017ucu,new} and we refer the reader to those papers for details. We will treat the semilocal network as a hybrid network made of local strings and global monopoles.  We will also assume low values of $\beta$, deep in the stability regime. \\

From a cosmological point of view, a crucial property of defect  networks is that they often show {\it scaling}: after some time, statistical properties of the network such as the typical inter-defect distance (or, if the defects have a spatial extension, their typical curvature radii) become a fixed fraction of the horizon size. This is true of cosmic strings and is also true of global monopoles so it is natural to ask whether a network of semilocal strings will also show scaling behaviour. This question has been analysed  in a series of papers with increasingly powerful numerical simulations \cite{Achucarro:1998ux,Achucarro:2005tu,Achucarro:2013mga,Lopez-Eiguren:2017ucu} that consistently showed  evidence of scaling behaviour. A first attempt at a (semi) analytical understanding was provided in \cite{Nunes:2011sf} in the form of two "velocity-one-scale" models,  with equations capturing the time evolution of the typical lengthscales and velocities of a network of semilocal strings.  In the next sections we build on these results to propose a new, improved, VOS model for semilocal networks together with some preliminary comparisons with numerical simulations.

\section{ A Velocity-dependent One-scale Model for semilocal strings} 

The velocity-dependent one-scale (VOS) model provides a quantitative and physically clear methodology to describe the evolution of topological defect networks, both in condensed matter and cosmological settings, and has been extensively compared to (and calibrated by) numerical simulations. It was first developed for cosmic strings \cite{VOS1}, and subsequently extended for several other defects. A recent review can be found in \cite{VOSbook}. Conceptually, it draws on Tom Kibble's insight \cite{Kibble} that defects' networks can be described by averaged quantities, for which evolution equations can be obtained starting from the defects' microscopic equation of motion. The term 'one-scale' expresses the assumption that the correlation length and defect curvature radius coincide with a characteristic length scale (or defect separation), henceforth denoted $L$. The VOS model extension of Kibble's one-scale model is based on the realization that a dynamical equation for the defect network's RMS\footnote{root-mean-square} velocity, henceforth denoted $v$, is crucial for any description that is able to model the evolution in various different settings, such as the radiation and matter eras in cosmology, and the transitions between them. \\

The VOS model therefore aims to provide a statistically averaged (macroscopic) description of the microphysical processes relevant for the dynamics of the defect networks. For some of these, the macroscopic description can be obtained directly from the microphysics in a mathematically rigorous way; an example is the effect of the cosmological expansion on the defects. For others, however, one is forced to introduce phenomenological parameters quantifying the relative importance of various terms, and these parameters cannot be calculated ab initio, but need to be determined by calibrating the model using numerical simulations; an example are the terms describing the various decay channels (or energy losses) for the network. For the purposes of the present work, the two key phenomenological parameters are a momentum parameter, $k$, related to the curvature of the defects, and an energy loss term, $c$. In the specific case of cosmic strings the latter parameter is often known as the loop chopping efficiency (loop production being thought to be the predominant decay channel for strings), but  analogous parameters exist for all other defects. \\

\bigskip
\subsection{Global monopoles}
A  VOS model for global monopoles was first developed in \cite{Martins} and revisited in \cite{Sousa}. With the model parameters as described above, and using the index $m$ in $L_m$ and $v_m$ to denote that these refer to monopoles,  we have the following dynamical equations
\bq
\dtot{v_m}{t} & = & \left(1-v_m^2\right)\left[\frac{k_m}{L_m}\left(\frac{d_H}{L_m}\right)^{\alpha} - \frac{v_m}{\ell_d}\right]\label{vosvnew}\,,\\
3\dtot{L_m}{t} & = & 3HL_m +\frac{L_m}{\ell_d}v_m^2+cv_m\label{vosLnew}\,.
\eq
where $v_m$ is the root-mean-square velocity of the monopoles and $L_m$ an overall length scale representing the distance between monopoles\footnote{In the simulations presented in the next section we will use a definition of $L_m$ based on the number density of monopoles in the box as  $L_m =$ (Total volume / Total number of  monopoles)$^{1/3}$.}. $H$ is the Hubble parameter. The parameter $c$ is the monopole analogue of the loop chopping efficiency for strings, and
$\ell_d$ is a damping length scale\footnote{The definition is  $( 1/ {\ell_d}) =2 H + (1/{l_f})$.  In the numerical simulations we will ignore particle scattering and take $l_f \rightarrow \infty$.} (including both the effects of Hubble damping and of friction due to particle scattering). \\

The first term in the right hand side of  eq.(\ref{vosvnew}) models the acceleration due to the surrounding monopoles and antimonopoles.  $d_H$ is the cosmological horizon, and $ \alpha$ is a parameter that describes the overall effect, on a given monopole, of all the other monopoles: $\alpha = 3/2$ corresponds to assuming that the force exerted on any given monopole by $N$ randomly distributed monopoles decreases as $F_{tot} \sim 1 /\sqrt N$. On the other hand, if we assume that the acceleration $a$ of a monopole is Brownian, so $a \sim \sqrt N$, we would have $\alpha = -3/2$. For networks of global monopoles, it was suggested in \cite{Sousa} that the second assumption gives a better description of the time evolution, though this needs to be confirmed by a robust statistical analysis. Here the situation is slightly different because the "monopoles" have a string attached to them that distorts the field profiles and the forces acting on the monopoles. We shall leave this parameter free and let the simulations determine the best fit. As it happens, the preferred value appears to be $\alpha=3/2$. \\
  
\subsection{Semilocal networks: previous attempts}

To first approximation, a VOS model for semilocal networks can be built by describing them as local strings with endpoints attached to global monopoles. Thus the VOS equations for monopoles in the previous sub-section should apply, with the caveat that in principle the velocity equation should include a term describing the force on the monopoles due to the strings. In practice, however, this force is subdominant when compared to the forces between the monopoles themselves---it is clear both from simulations and analytic modeling that the semilocal network dynamics is controlled by the monopoles. \\

On the other hand, a semilocal network is made of string segments, each of which will have a length $l_s$ and velocity $v_s$ \footnote{the index $s$ denotes that $l_s$, $v_s$ refer to the strings}. In principle the evolution of closed string segments (e.g., loops) is well-known for standard topological strings. In the semilocal case, there is an additional dynamical mechanism which will affect the evolution of the segments: the gradient energy acts as a long-range force between the monopoles, which can determine whether string segments shrink or grow. In \cite{Nunes:2011sf}, two models were proposed for the evolution of $l_s$ and $v_s$, based on different (but reasonable) simplifying assumptions. The first stems from the hypothesis that the evolution of each segment is the result of the competition between the timescales for segment annihilation, and monopole-antimonopole annihilation: if the former is shorter the segment with shrink and decay (which is expected to be the most likely outcome for small segments), while if the latter is shorter two segments will merge and a larger one will be formed (which is expected to be the most likely outcome for large segments). This model also assumes luminal motion  ($v_m \sim 1$) for the monopoles at the ends of the segments. The second model stems from a balance equation for a network of segments, which is assumed to be Brownian. \\

\begin{figure}[!h]
\centering\includegraphics[width=4in]{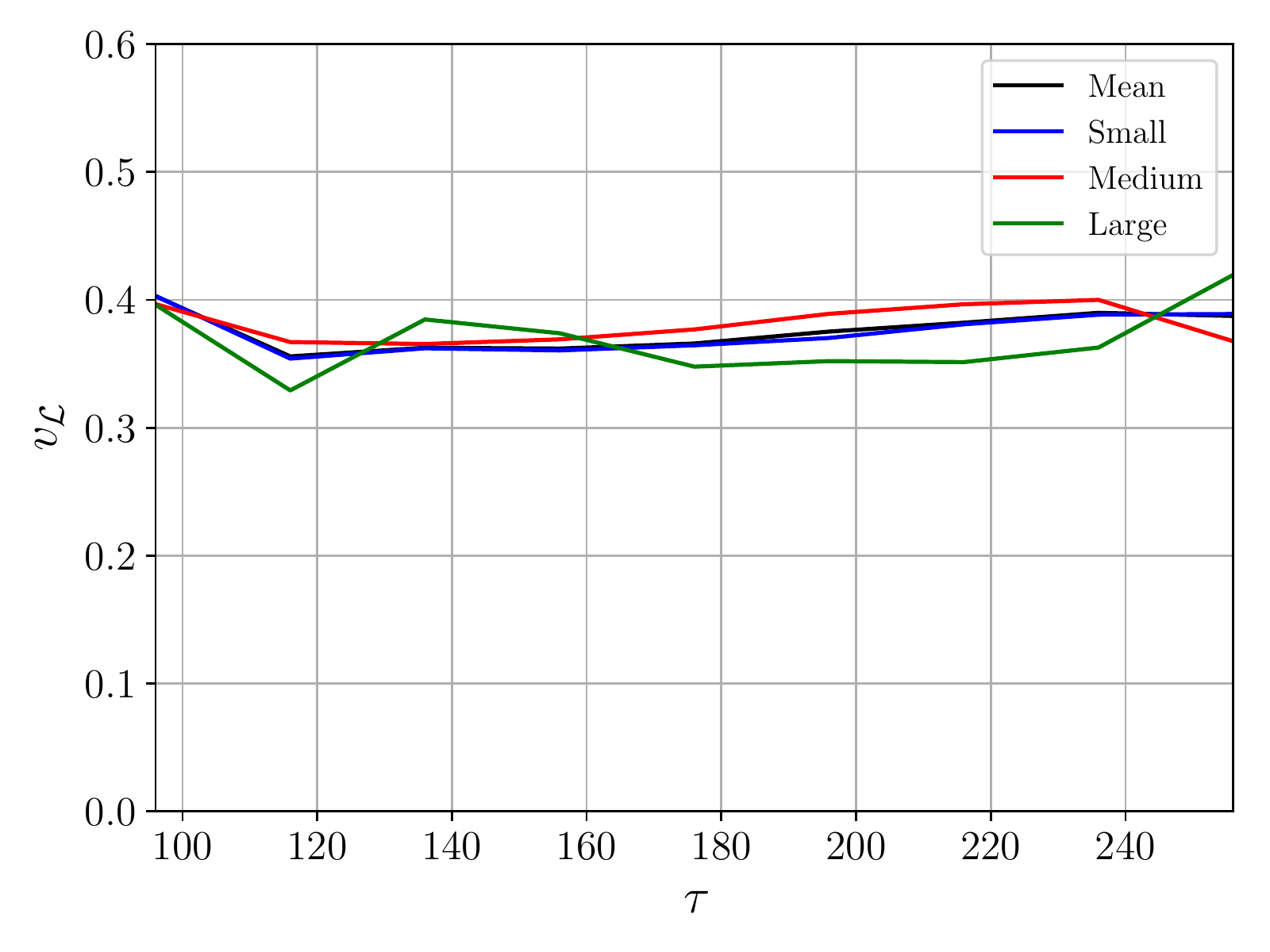}
\caption{The average velocity of string segments, showing no difference between large, medium and small segments. 
In this plot the small segments are defined as segments with lengths smaller than $ls_{min}+(ls_{max}-ls_{min})/3$, where $ls_{min}$ is the smallest segment in each time-step and $ls_{max}$ is the longest one. Medium segments have lengths between  $ls_{min}+(ls_{max}-ls_{min})/3$ and  $ls_{min}+2*(ls_{max}-ls_{min})/3$ and long segments are segments longer than  $ls_{min}+2*(ls_{max}-ls_{min})/3$. Mean means the average over all segments in the box. The velocities in this plot are computed using the seven simulations for $\beta=0.15$ radiation era.}
\label{fig_segmentvelocity}
\end{figure}

While both models give a good description of the segments' evolution on short timescales, they fail to describe the long-term behaviour. A possible reason is that both models assume that smaller segments would move faster than longer ones. A priori this is a natural assumption: specifically, it is motivated by the topological string case, where smaller loops are indeed faster than longer ones. However, in the semilocal case this effect is not borne out by the simulations, see figure \ref{fig_segmentvelocity}. It shows that the average speed of the semilocal segments is largely independent of their length. \\

\subsection{Semilocal networks: a new proposal}

Our new proposal for the analytic description of semilocal string networks extends our earlier VOS model, based on the hypothesis that the semilocal string dynamics is determined by the competition between the segment annihilation and monopole annihilation timescales, by relaxing the assumption that the monopole velocity be luminal, $v_m \sim 1$. Instead, the model evolves the monopole RMS velocity using equation (\ref{vosvnew}). The evolution of the length of any given segment is given by  
\be
\frac{dl_{s}}{dt}=Hl_{s}-v_{s}^{2}\frac{l_{s}}{l_{d}}+\sigma\left(1-\frac{L_m}{l_{s}}\right)v_{m}^{2} \label{dlsdt} \,,
\ee
where the third term describes the competition between string segment and monopole-antimonopole annihilation. Note that this term crucially depends on the monopole velocity $v_m$ and there is a free parameter, $\sigma$, to be fitted by comparison to numerical simulations. The characteristic lengthscale for monopoles, $L_m$, can be read from the simulations or implicitly given by the string segment evolution (as the number of monopoles is twice the number of string segments). Finally, the model drops the previous assumption of an evolving velocity distribution for string segments, and instead keeps the segment velocities, $v_s$, constant as seen in the simulations. \\

An important difference between semilocal and topological string networks is that, in the former, there is no single characteristic length scale for the string network. Instead, there is a distribution of segment lengths in the network, and each segment's length evolves according to (\ref{dlsdt}). Scaling then corresponds to the total string network length evolving linearly in time, and the segment lengths reaching a steady state distribution. Note that the total number of segments is not conserved. When a segment shrinks to zero length it is removed from the network and, correspondingly, the total number of monopoles reduces by 2. In the next section we present our preliminary results comparing our VOS model to field theory simulations of semilocal string networks, supporting scaling.   \\

\section{Numerical evolution of the semilocal network}

We are interested in simulating the semilocal string network evolution in the early universe, in particular during the radiation and matter domination eras.  In order to do that we solve the equations of motion  derived from 
the semilocal model 
(\ref{SLaction}).  We work in comoving coordinates and conformal time $\tau$ in a spatially  flat Friedmann-Robertson-Walker space-time:
\be
g_{\mu\nu}=a(\tau)^2\eta_{\mu\nu}\,,
\ee
 where $\eta_{\mu\nu}={\rm diag}(-1,1,1,1)$ is the Minkowski metric  
 and $a(\tau)$ is the cosmic scale factor.   The scale factors we will consider are  those corresponding to a radiation dominated universe ($a\propto \tau$) and to a matter dominated universe  ($a\propto \tau^2$).
In what follows $\dot{} \equiv d / d\tau = a ~ d/dt $. \\

Choosing  the temporal gauge $(A_0=0)$, the field equations read
\begin{eqnarray}
\dot\Pi + 2\frac{\dot a }{a}  \Pi -\mathbf{D}^2\Phi + a^2\beta (|\Phi|^2 -1)\Phi &=& 0, \nonumber\\
\partial^\mu   F_{\mu\nu}  - ia^2(\Phi^{\dagger}D_\nu\Phi - D_\nu\Phi^{\dagger}\Phi) &=& 0,
\label{eom}
\end{eqnarray}
where  and $\Pi=\dot\Phi$, together with Gauss's law
\begin{equation}
\partial_i  F_{0i} =   i a^2 [ \Pi^\dagger \Phi - \Phi^\dagger \Pi ] .
\end{equation} 

Our aim is to simulate a network of defects in an expanding background. Specifically, in a radiation dominated universe ($a\propto \tau$) and in a matter dominated universe  ($a\propto \tau^2$). However, this 
presents us with another difficulty:  the physical size of the defects is fixed throughout the 
simulation, but  the size of the simulation box is growing with time. Equivalently, viewed in comoving coordinates, the box 
size is constant, but the comoving size of the defects shrinks.  \\

One solution to this problem is to use an algorithm presented by  Press-Ryden-Spergel (PRS)  \cite{Press:1989yh} for global defects, and extended in \cite{Bevis:2006mj} for local ones.  
With this procedure the defect cores are made to grow artificially, in order not to lose them during simulation. The (continuum) equations of motion  (\ref{eom})  get modified to:
\begin{eqnarray}
\dot\Pi + 2\frac{\dot a }{a} \Pi -\mathbf{D}^2\Phi + a^{2s}\beta (|\Phi|^2 -1)\Phi &=& 0, \nonumber\\
\partial^\mu \left(a^{2(1-s)}  F_{\mu\nu} \right) - ia^2(\Phi^{\dagger}D_\nu\Phi - D_\nu\Phi^{\dagger}\Phi) &=& 0 .
\label{eom2}
\end{eqnarray}

The parameter $s$ interpolates between the true equations of motion (for $s=1$)  and an unphysical situation where   comoving core size of the defects is constant ($s=0$). 
As discussed in \cite{Lopez-Eiguren:2016jsy} (and references therein) the $s=0$ case is a good approximation to the true $s=1$ evolution in the semilocal case.\\

Using a discretized version of the PRS modified equation of motion, the system was simulated in $1024^3$ cubic lattices with periodic boundary conditions for radiation and matter domination. The values of $\beta$ chosen for the simulations were $\beta<<1$ where a considerable number of defects is known to form, in order to have enough string segments to later use in the comparison with VOS models.  We performed seven simulations  for each $\beta=0.04,0.09,0.15,0.20,0.25,0.30, 0.35$. The initial conditions chosen had the gauge field, gauge field velocities and scalar field velocities set to zero; and the scalar fields were chosen to lie in the vacuum manifold, but with randomly chosen orientations. An initial short period of forced damping was used to give the system an opportunity to relax to the scaling regime. Then, the simulation was run until half light crossing time, an upper bound in time set by the periodic boundary conditions. \\

Here we are interested in comparing this VOS model with the simulations. In order to do that we have developed algorithms to estimate the number of string segments and their lengths, the number of monopoles, and the velocities of segments and monopoles. These can be found in a series of previous papers (for a detailed discussion see \cite{Lopez-Eiguren:2017ucu}). Here we summarize the main features and refer the reader to the original work for details.  \\

When simulating networks of topological defects, one can make use of their topological nature  to detect them in the simulated box. However, as mentioned earlier,  semilocal strings are not topological, and it makes it much more difficult to numerically detect their location in a simulated lattice. One criterion used in the literature \cite{Achucarro:1998ux,Achucarro:2005tu,Nunes:2011sf,Achucarro:2013mga} was to identify points in the lattice with a certain concentration of magnetic flux. This did ensure that those points were part of semilocal strings, and did allow us to obtain the number  densities of strings and estimate their lengths. This was done by grouping points of high magnetic flux by proximity, and thus this was an estimator of  the "volume" of the strings.  The number density of strings is also a good estimator for the number density of monopoles. \\

In \cite{Lopez-Eiguren:2017ucu} we showed that this estimator seriously understimated the string length, and another series of estimators were introduced. One of them used the windings of the scalar fields (of $\Phi_1$, $\Phi_2$ or both fields together) to obtain a one-dimensional characterization of the network. Having a one dimensional estimator was invaluable, because on top of allowing for a more accurate description of the length of the strings, it provided the means of obtaining an estimation of the velocities of the segments (transverse velocity of the strings) for the first time, and thus enabling the comparison with VOS models. \\

Another improvement included in \cite{Lopez-Eiguren:2017ucu}  was the use of the monopole-like nature of the string ends. The whole simulated volume is analysed point by point to check whether there exist points which have "monopole" topological charge, and those points were identified as monopoles or anti-monopoles. Once again, having the one-point position of the monopoles enabled us to obtain monopole velocities. Moreover, this method of monopole detection is independent of the string-segment detection, and thus both can be combined and compared to check the robustness of the methods. All these estimators have some important caveats, and we refer the interested reader to  \cite{Lopez-Eiguren:2017ucu}) for a more in-depth description/discussion. \\

Using the string lengths, string velocities and monopole/string end velocities we can analyse the newly introduced analytical model that describes the evolution of the semilocal segments. In order to do that we use the data coming from the simulations, where we extracted data at nine equally spaced time-steps. The data from the first time-step is used to seed the analytical model and then the equations are evolved. We extract data from the analytical models at eight equally spaced time-steps. These time-steps coincide with the simulation time-steps. Therefore, we have two sets of segment distributions at eight time-steps (eight, rather than nine, because the first output time-step from the simulations has been used to seed the analytical models). \\

\begin{figure}[!h]
\centering\includegraphics[width=4.0in]{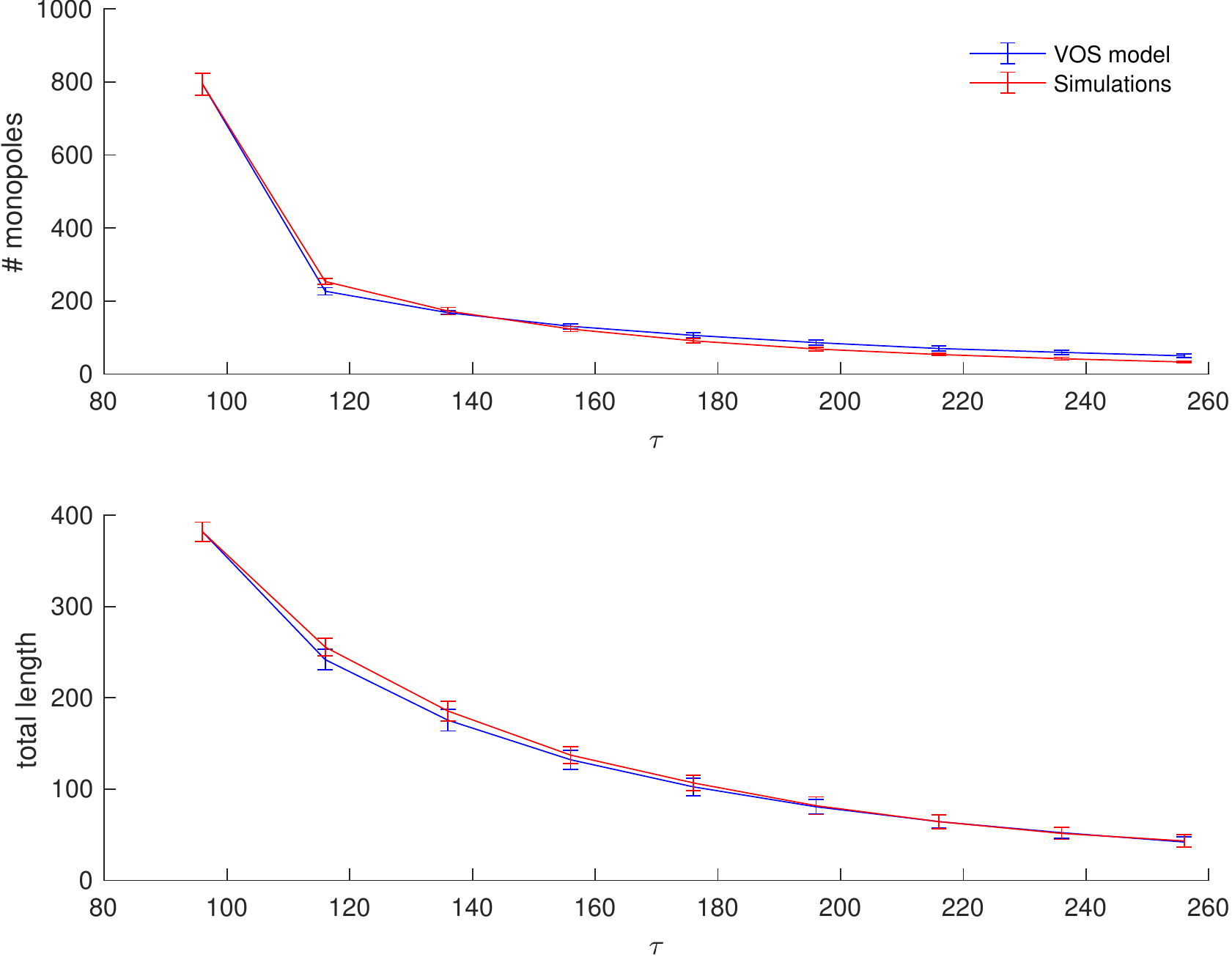}
\caption{Evolution of the total number of monopoles and the total segment length computed from the field theoretical simulations (red) and from the VOS model (blue).  The  graphs show the average over seven runs for 
$\beta=0.15$ in the radiation era.  To obtain the VOS data, the VOS equations (\ref{vosvnew}) and (\ref{dlsdt}) with parameter values $\sigma=2.7$ and $k_m=1.6$ were solved seven times, each of them taking the initial conditions from one of the seven simulations, and the results were averaged.
}
\label{fig_res}
\end{figure}

\begin{figure}[!h]
\centering\includegraphics[width=5.0in]{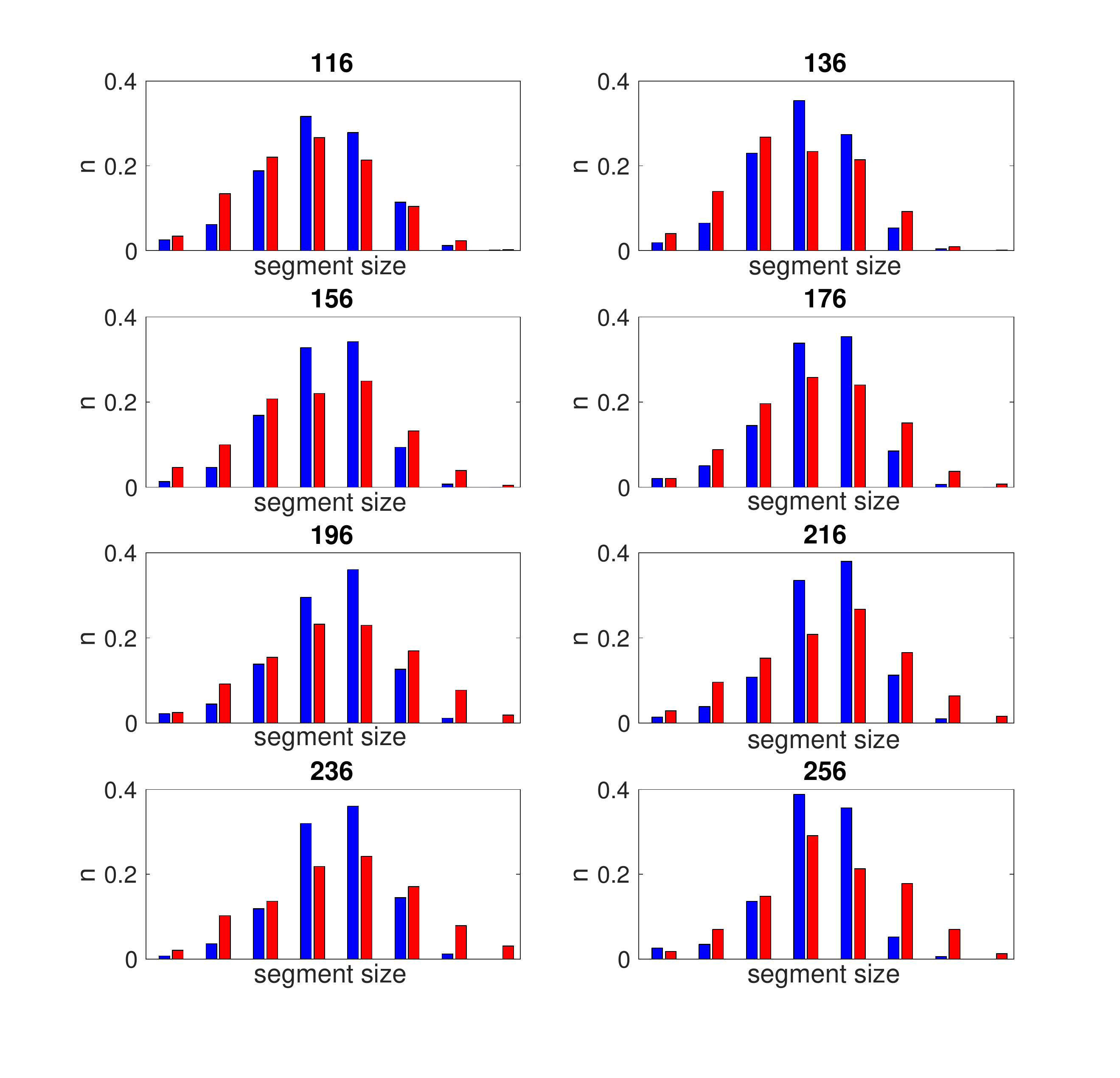}
\caption{The distribution of segment lengths computed from the field theoretical simulations (red) and from the VOS model (blue).  The figure shows histograms at various, equally spaced,  timesteps from t=116 until t=256,. At each timestep we identify the longest and the shortest segment and all the segments are binned into eight equal-length bins. $n$ is the number of segments in the bin normalized to the total number of segments.  The  red histograms show the average over seven runs for 
$\beta=0.15$ in the radiation era.  To obtain the VOS data, the VOS equations (\ref{vosvnew}) and (\ref{dlsdt}) with parameter values $\sigma=2.7$ and $k_m=1.6$ were solved seven times, each of them taking the initial conditions from one of the seven simulations, and the results were averaged. }
\label{VOS-histograms}
\end{figure}

In order to perform a $\chi^2$ analysis to find the parameter values for which the VOS model describes better the evolution of the system, we bin the segments in each time-step depending on their lengths. The $\chi^2$ is computed comparing the number of segments in each bin and in each time step. Our preliminary analysis of the new VOS model shows that the evolution of the total string length and the total number of segments is well described, see Fig~\ref{fig_res}. Moreover, the monopole velocities obtained from the VOS model agree with the velocities obtained from the simulations. For example, at radiation era with $\beta=0.15$ the monopole velocity obtained from the simulations is $v_{\mathcal{M}}=0.600 \pm 0.012$ and the one obtained from the analytical model is $v_{\mathcal{M}}=0.62 \pm 0.02$.  Finally, preliminary results on the distribution of segment lengths as time progresses are shown in Fig~\ref{VOS-histograms}.  The VOS model tends to predict more medium-length segments and fewer long segments than what is observed in the simulations, but in general the distributions seem to agree remarkably  well.
\\

This preliminary analysis shows that the new model is a good candidate. However, a more detailed analysis will be necessary to certify its validity. This we leave for future work \cite{new}.  \\



\maketitle

\section{Conclusions}

Semilocal strings are  non-topological defects closely related to Abelian Higgs (topological) strings and also to the electroweak strings and dumbbells of the Standard Model. Infinite, straight semilocal strings are stable if the mass of the scalar fluctuations is much smaller than the mass of the vector fluctuations, a regime that would be called "deep type I" in ordinary superconductors. In this stability regime ($\beta << 1$), a cosmological phase transition would lead to a surviving network composed of closed loops and string segments whose ends have long-range interactions. As a result, the segments can either shrink or grow to join other segments and form longer ones or closed loops. There is numerical evidence that such networks can reach scaling. However, an analytic description of the overall evolution of the network has so far remained elusive.\\

We have presented a "thermodynamical" (velocity-one-scale) model for the network, given by eqs. (\ref{vosvnew}),(\ref{dlsdt}). 
We simultaneously model an overall lengthscale and RMS velocity for the monopoles, and the length of each string segment.
The equation for the segment lengths was first derived in ref. \cite{Nunes:2011sf} and the equation for the monopoles in refs. \cite{Martins, Sousa}. We presented preliminary evidence that the network evolution is well described by this model.  \\

An important departure from  previous proposals is that we take the velocity of semilocal segments to be independent of their length. This is a remarkable difference between the semilocal and the Abelian Higgs case, where smaller loops move faster than longer ones. \\

Our results would appear to confirm that the network evolution is dominated by the dynamics of the ``global monopoles" at the ends of the segments, and by their long-range interactions. However, the effect of the strings is seen in the parameters that appear in the equation for the global monopole RMS velocity, eq. (\ref{vosvnew}), where the multi-monopole effects are somewhat different from the case of pure global monopoles.  \\

\enlargethispage{20pt}


\dataccess{The code and the data from the simulations are publicly available at the following URL: https://bitbucket.org/Lopez-Eiguren/semilocal/src/master/}

\aucontribute{This work is the result of a collaborative effort with the numerical work performed mainly by Asier L\'opez-Eiguren and Jon Urrestilla, VOS modelling mainly by C.J.A.P Martins, A. Avgoustidis and  A. Ach\'ucarro.  All authors contributed to the writing of the paper. They have read and approved the manuscript.}

\competing{The authors declare that they have no competing interests.}

\funding{A. Ach\'ucarro is partially supported by the Netherlands'  Organization for Fundamental Research in Matter (FOM) and by the Delta-ITP gravity program (NWO/OCW).  A. Ach\'ucarro and J. Urrestilla 
acknowledge support from the Basque Government/Eusko Jaurlaritza (IT- 979-16), and by the Spanish Ministry MINECO FPA2015-64041-C2-1P (MINECO/FEDER) and PGC2018-094626-B-C21 (MCIU/AEl/FEDER,UE). 
A. Avgoustidis was supported by STFC grant ST/P000703/1. 
A. L\'opez Eiguren   is supported by the Academy of Finland grant 286769. 
The work of C.J. Martins was financed by FEDER---Fundo Europeu de Desenvolvimento Regional funds through the COMPETE 2020---Operational Programme for Competitiveness and Internationalisation (POCI), and by Portuguese funds through FCT - Funda\c c\~ao para a Ci\^encia e a Tecnologia in the framework of the project POCI-01-0145-FEDER-028987. \\
Part of this work was undertaken on the COSMOS Shared Memory system at DAMTP, University of Cambridge operated on behalf of the STFC DiRAC HPC Facility. This equipment is funded by BIS National E-infrastructure capital grant ST/J005673/1 and STFC grants ST/H008586/1, ST/K00333X/1. It has been possible also thanks to the computing infrastructure of the i2Basque academic network. \\ }




\end{document}